\documentclass[fleqn,twoside]{article}
\usepackage{espcrc2}
 
\usepackage{epsfig}
\usepackage[figuresright]{rotating}
 
\newcommand{\AmS}{{\protect\the\textfont2
  A\kern-.1667em\lower.5ex\hbox{M}\kern-.125emS}}
 
 \def\xgo{x_\gamma^{\rm obs}}
 
\title{Open charm production at HERA}
 
\author{U. Karshon\address[MCSD]{Weizmann Institute of Science, Rehovot, 
        Israel \\                 
       On behalf of the H1 and ZEUS Collaborations}%
        \thanks{Supported by the Israel Science Foundation and the 
                U.S.-Israel Binational Science Foundation.}  }        
       
\begin{document}
 
\begin{abstract}
 
Inclusive charm meson production cross sections in the deep inelastic scattering       and
photoproduction       regimes are compared with QCD leading and next-to-leading order
(NLO) calculations. The NLO predictions are significantly below the data in some parts
of the measured kinematic range. Angular distributions of dijet events with charm
show clear evidence for the existence of charm originating from the incoming photon.
The charm fragmentation function is measured for the first time at HERA. Various
fragmentation ratios and the fragmentation fractions of the low-lying charm hadrons
are determined and compared to previous $e^+ e^-$ results.
 
\vspace{1pc}
\end{abstract}
 
\maketitle
 
\section{Introduction}
 
During 1996-2000 HERA collided electrons or positrons ($E_e$=27.6~GeV) with protons
($E_p$ between 820 and 920~GeV). Open charm ($c$) production, which has been extensively studied,
occurs in few steps: a) Hard process, e.g. boson-gluon fusion (BGF), where a photon ($\gamma$) or Z
boson emitted from the incoming electron fuses with a gluon ($g$) from the proton, producing
a $c\bar c$ pair; b) Initial/final state parton shower development; c) Fragmentation of
a final-state parton into a hadron. Two kinematic regions have been explored: 
1) Deep inelastic scattering (DIS) with photon virtuality                                       
                                   $Q^2 > 1~GeV^2$, where the scattered electron is visible
in the main detector; 2) Photoproduction (PHP) with $<Q^2>\approx~3\cdot~10^{-4}~GeV^2$,
                where the virtual photon radiated from the incoming electron is quasi-real.
 
The large mass of the $c$ quark provides a ``hard" scale needed for the comparison of data
to QCD predictions. In leading order (LO) QCD, two types of processes are responsible for    
charm PHP: Direct photon processes, where the photon interacts as a point-like particle
and resolved photon processes, where the photon acts as a source of partons. BGF         
       is the dominant direct photon process. Charm quarks present in the parton distributions
of the photon lead to LO resolved charm excitation processes like $cg\to cg$.
 
Various                        NLO  calculations exist for comparison with charm HERA
data: 1) fixed-order (FO) approach~\cite{FO}, 
        where only u,d,s are active flavours in the photon and proton; 
2) resummed (RS) approach~\cite{kniehl,cacciari},         
                   where a ``massless" $c$ quark is also an active flavour;
3) matched (FONLL) calculation~\cite{FONLL}, which incorporates mass effects up to NLO
and the resummation of $p_T$ logarithms up to next-to-leading logarithm (NLL) level. All these
 approaches are based on the DGLAP evolution~\cite{DGLAP}. The CASCADE~\cite{Jung}
Monte Carlo (MC), based on the CCFM evolution~\cite{CCFM}, is a more recent tool to compare 
with the data.
 
Charm can be tagged by reconstructing charm mesons via decays
                                such as $D^{*+}\to~D^0\pi^+\to~(K^-\pi^+)\pi^+$,
$D^+\to~K^-\pi^+\pi^+$,
$D^0\to~K^-\pi^+$, 
$D_s^+\to~\phi\pi^+\to~(K^- K^+)\pi^+$, where charge conjugate states are included, or by 
measuring semileptonic electrons or muons from charm. Reconstructing charm hadrons via their
decay vertices with a silicon tracker significantly reduces the background.
 
\section{Inclusive charm meson production}
 
Inclusive production of the charm mesons $D^{*+}$, $D^0$, $D_s$ and $D^+$ in 
    DIS  has been studied~\cite{H1DIS} with the H1 detector,
                                        using a sample corresponding to an integrated 
luminosity of $48~pb^{-1}$.
 A silicon tracker separates the     production and decay vertices
of the pseudoscalar mesons. Differential cross sections have been measured for the $D^+$ and
$D^0$ mesons                                                                  
             and compared with the AROMA~\cite{aroma} LO MC predictions  
             in the kinematic range $2~<~Q^2~<~100~GeV^2$, $0.05~<~y~<~0.7$,
$p_t(D)~>~2.5~GeV$ and $|\eta(D)|~< 1.5$, where $y$ is the electron inelasticity and $p_t$
and $\eta$ are, respectively, the transverse momentum and pseudorapidity of the $D$ meson.
$\eta$ is defined to be positive in the proton beam direction.
Results for the $D^+$ channel are shown in Fig.{1}. The lower shaded bands indicate the AROMA
beauty contribution, scaled by the excess factor of data over MC, as found in  the
H1 publication~\cite{H1b}. The distributions are well described by the LO QCD
simulation, both in normalisation and shape.
 
       \begin{figure}[h]
 \vspace{-0.3cm}
          \begin{center}
          \psfig{figure=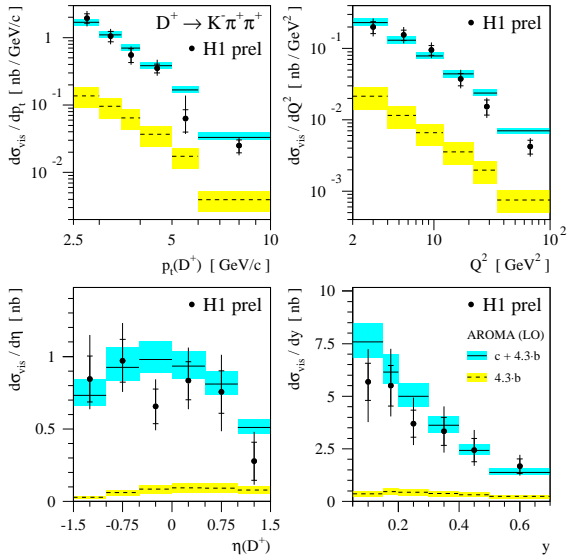,height=7.5cm}
          \end{center}          
 \vspace{-0.9cm}
\caption{Differential cross sections for $D^+$ as a function of the $D^+$ (left) and
event (right) variables. The AROMA MC predictions are given by the shaded bands.}
\label{fig:fig1}
       \end{figure} 
 
       \begin{figure}[h]
 \vspace{-0.3cm}
          \begin{center}
          \psfig{figure=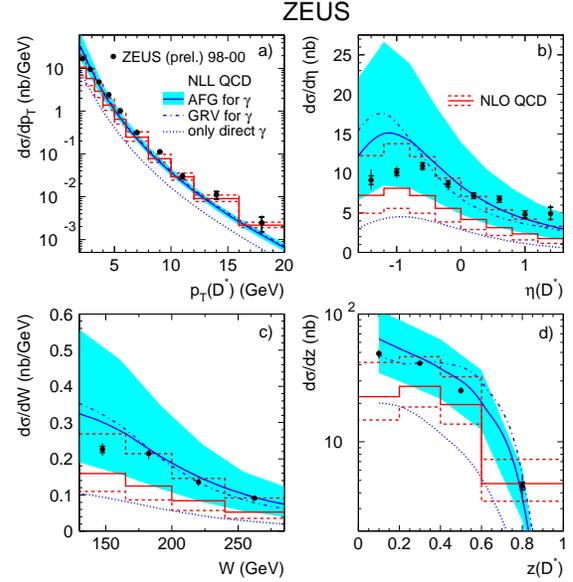,height=8.2cm}
          \end{center}          
 \vspace{-0.9cm}
\caption{Differential cross sections for inclusive $D^*$ production as a function of
 $p_T$, $\eta$, $W$ and $z$. FO predictions with nominal parameters are given by solid
histograms. Upper (lower) dashed histograms correspond to upper (lower) bounds of the predictions.
NLL predictions are shown by solid curves and shaded bands (dash-dotted lines) for the AFG (GRV)
      photon structure function
parametrisation.     Direct photon NLL predictions are given by the dotted lines. }
\label{fig:fig2}
 \vspace{-0.4cm}
       \end{figure} 
 
Inclusive PHP of $D^{*+}$ mesons has been measured~\cite{ichep786}                         
                                                   with the ZEUS detector in the kinematic 
region $Q^2~<~1~GeV^2$,                photon-proton
centre-of-mass energies $130~<~W~<~285~GeV$,                                      
$1.9~<~p_t(D^*)~<~20~GeV$ and $|\eta(D^*)|~< 1.6$,
using an integrated luminosity of $79~pb^{-1}$.
The measured differential cross sections were compared with the  NLO 
predictions FO~\cite{FO}, RS~\cite{kniehl} and FONLL~\cite{FONLL}. In
fig.{2} the distributions $d\sigma /dp_T$, $d\sigma /d\eta$, $d\sigma /dW$ and
$d\sigma /dz$ are compared with the FO and RS calculations,                               
               where $z(D^*)$  is  the fraction of the photon energy carried by the $D^*$ 
meson in the proton rest frame. Theoretical uncertainties  obtained by varying the charm mass
and renormalisation scale     are large, in particular for the NLL predictions.       
The central FO predictions are below the data, mainly for $\eta > 0$
and low $z$. The NLL calculations are closer to the data. In particular NLL is better than
FO for $d\sigma /dz$ and for $\eta > 0$.
 A significant resolved contribution is required in the NLL predictions, which show some
sensitivity to the photon structure function parametrisation.
 
The precise ZEUS data enable measurements of double differential cross sections. In Fig.{3} the
$\eta$ distributions for four regions in $p_T$ are compared with FO and FONLL predictions.
The data is close to the upper band of the predictions and is significantly above the FO 
and FONLL calculations at medium $p_T$ and positive $\eta$, where even the upper bounds are
below the data. FONLL predictions are close to the FO ones for low $p_T$, but below FO
for large $p_T$, where FONLL should do better.                                              
 
       \begin{figure}[h]
 \vspace{-0.3cm}
          \begin{center}
          \psfig{figure=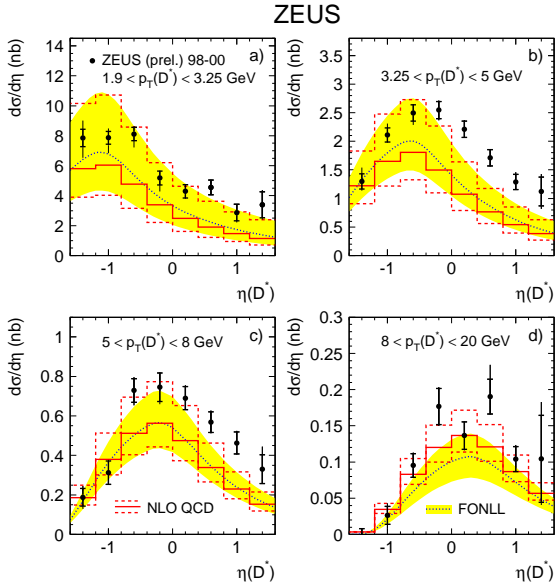,height=8.2cm}
          \end{center}          
 \vspace{-0.9cm}
\caption{Differential cross sections $d\sigma /d\eta$ for inclusive $D^*$ production
for four $p_T$ regions. FO predictions with nominal parameters are given by solid
histograms. Upper (lower) dashed histograms correspond to upper (lower) bounds of the predictions.
FONLL predictions are shown as dotted curves with uncertainties given by the shaded bands. }
\label{fig:fig3}
 \vspace{-0.5cm}
       \end{figure}

       \begin{figure}[h]
 \vspace{-1.1cm}
          \begin{center}
 
 \hspace*{-1.5cm}\psfig{figure=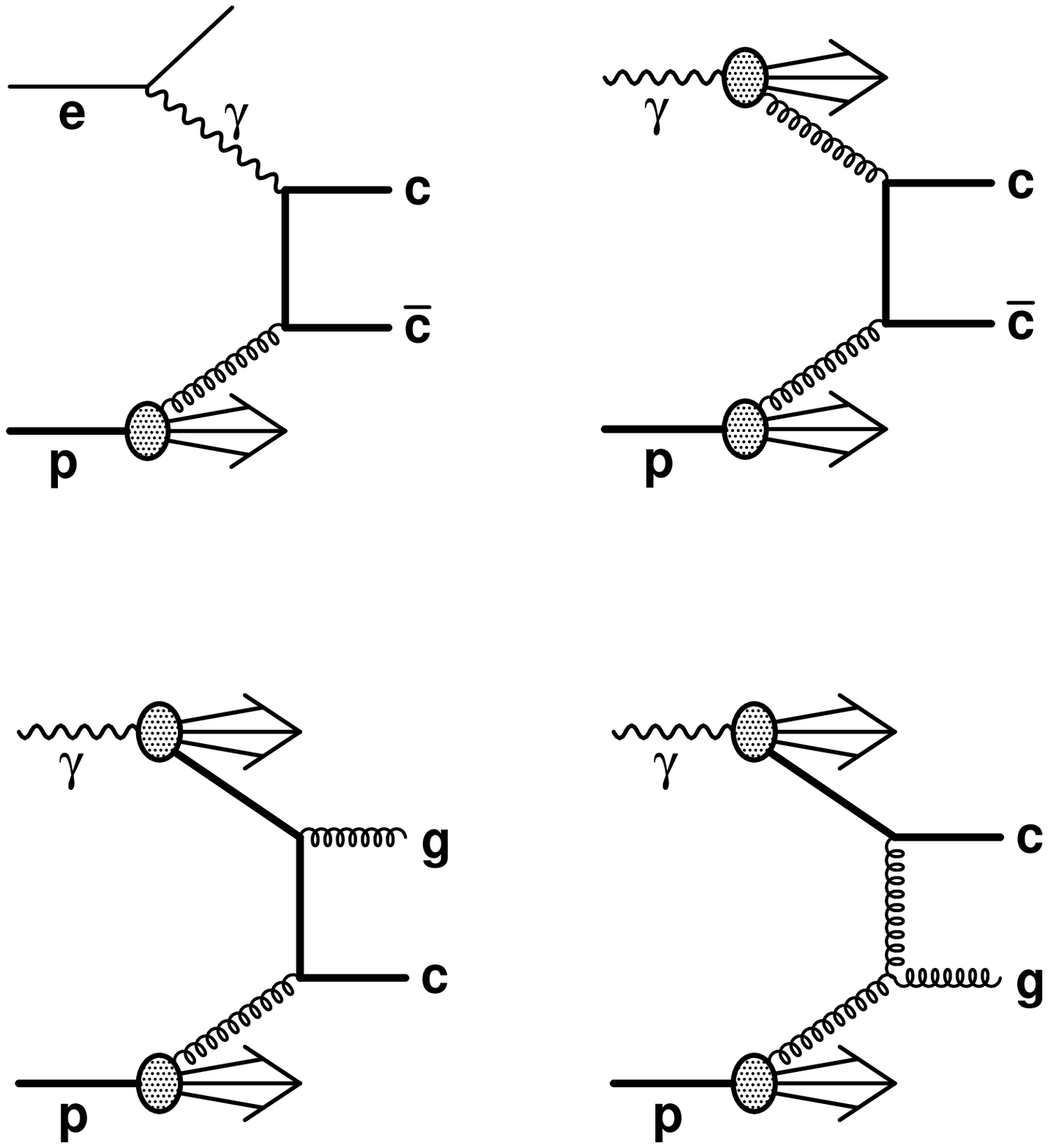,height=11.0cm}
 
 \vspace*{- 3.0cm}
 
 \hspace*{-1.0cm}\psfig{figure=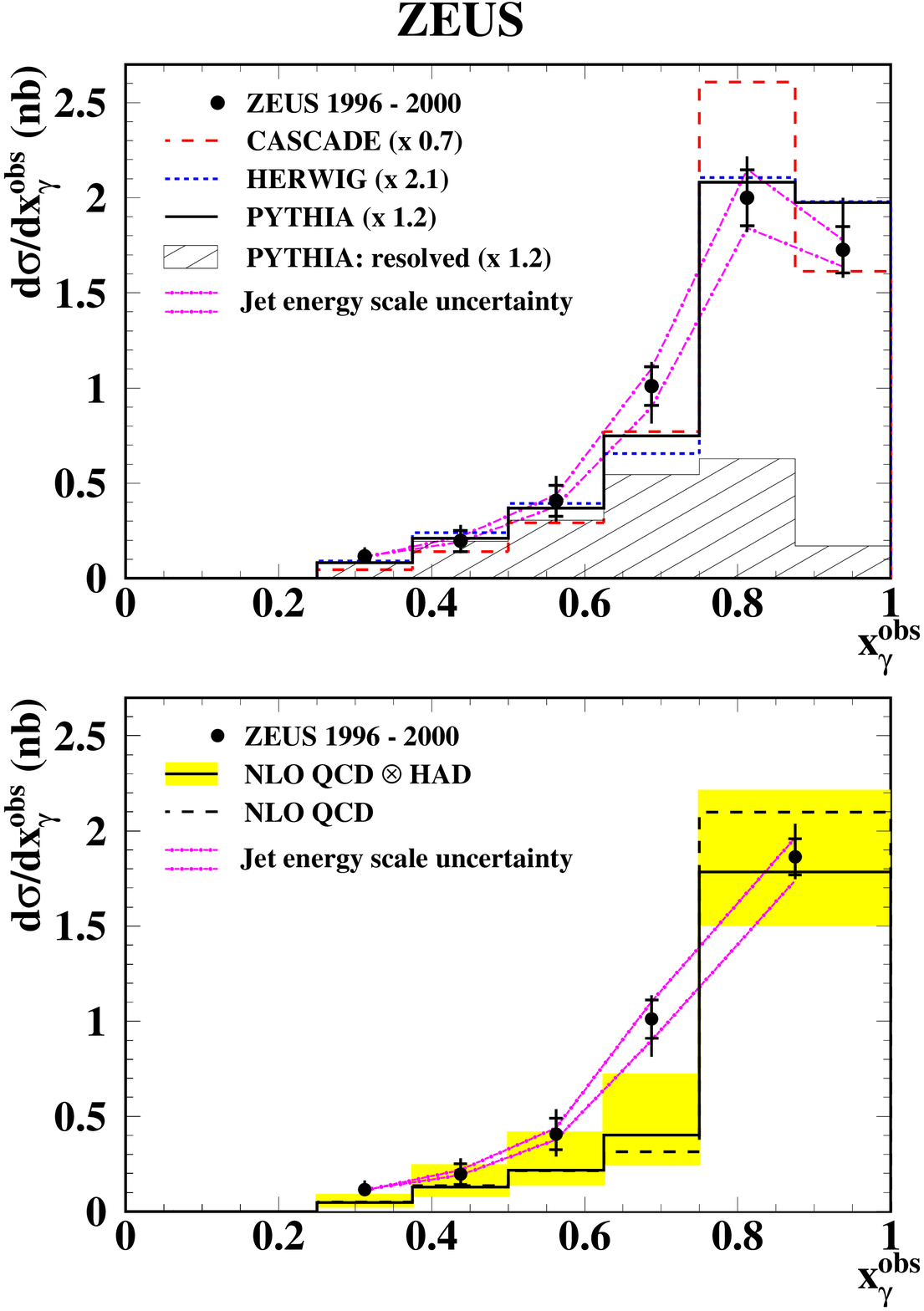,height=8.0cm}
 
          \end{center}          
 \vspace{-1.1cm}
\caption{Upper part: LO QCD charm-production diagrams. Direct photon, BGF ($\gamma~g\to~c\bar c$); 
resolved photon ($g~g\to~c\bar c$); resolved-photon charm excitation      
                ($c~g\to~g~c$, $c$ in proton hemisphere);                
                             resolved-photon charm excitation       
               ($c~g\to~c~g$, $c$ in photon hemisphere).                
Lower part:  Differential cross-section $d\sigma /d\xgo$ for the data compared     
with MC simulations (upper plot) and NLO FO predictions (lower plot). Each MC distribution
is normalised to the data, as indicated in the brackets. The NLO uncertainty after 
hadronisation is given by the shaded band. } 
\label{fig:fig4}
       \end{figure} 
 
\section{Charm dijet angular distributions}
 
Charm-dijet PHP events enable the study of the photon structure, in particular it's charm content.
Inclusive PHP of $D^*$-dijet events has been measured~\cite{cost} 
                                                         with the ZEUS detector in the kinematic 
region $Q^2~<~1~GeV^2$,
 $130~<~W~<~280~GeV$, $E_T^{jet}~>~5~GeV$, $|\eta^{jet}|~<~2.4$, $M_{jj}~>~18~GeV$,
$|\bar\eta|~<~0.7$, $p_t(D^*)~>~3~GeV$ and $|\eta(D^*)|~< 1.5$,
using an integrated luminosity of $120~pb^{-1}$.
Here $M_{jj}$ is the jet-jet effective mass and $\bar\eta$ is the average pseudorapidity of the
two jets. The fraction of the photon momentum producing the two jets is given by:
$x_\gamma^{\rm OBS} = \frac{\Sigma_{\rm jets} E_T e^{-\eta}}{2yE_e}$. This quantity is close
to 1 for     direct photons     and is less than 1 for resolved photons. A sample of direct-
(resolved-)enriched events is defined as   $\xgo~>~0.75~(~<~0.75)$.
 
The differential cross section $d\sigma /d\xgo$ is compared in                   
fig.{4} with LO predictions of the MC programs PYTHIA~\cite{PYTHIA}, 
    HERWIG~\cite{HERWIG} and CASCADE~\cite{Jung}  and
with NLO FO predictions~\cite{FO}. A significant contribution ($\approx 40\%$) arises from resolved
photons. There is a good agreement in shape between the data and the LO MCs, except that
CASCADE is too high for the high $\xgo$ region. The low $\xgo$ tail of the NLO prediction
is below the data.

The angle between the jet-jet axis and the beam axis in the dijet rest frame can be
approximated by
       $\cos\theta^* = \tanh \frac{\eta^{jet1} - \eta^{jet2}}{2}$.  In QCD
the differential cross section $d\sigma /d|\cos\theta^* |$ is sensitive to the spin of
the propagator in the hard subprocess. Whenever the propagator 
in the LO diagram is a quark, the cross section rises slowly as
                       $(1 - |\cos\theta^*|)^{-1}$, and when it is a gluon,            
 it rises steeply as
                       $(1~-~|\cos\theta^*|)^{-2}$.

Fig.~{5} shows 
        $d\sigma /d|\cos\theta^*|$ separately for the resolved-enriched ($\xgo~<~0.75$)
        and           direct-enriched ($\xgo~>~0.75$) samples.                     
  The LO MCs PYTHIA and HERWIG describe the shapes of both data samples.
The direct sample rises slowly with $|\cos\theta^*|$, consistent with the q-exchange BGF    
up-left diagram of fig.{4}.
The resolved sample rises strongly with $|\cos\theta^*|$, indicating a g-exchange signature.
Consequently, the LO subprocess $gg\to c\bar c$ (up-right diagram of fig.{4}) cannot dominate
the resolved photon charm-dijet process. This suggests the dominance of the LO charm-excitation
process $c g\to c g$ (down-right diagram of fig.{4}).

       \begin{figure}[h]
 \vspace{-0.6cm}
          \begin{center}
          \psfig{figure=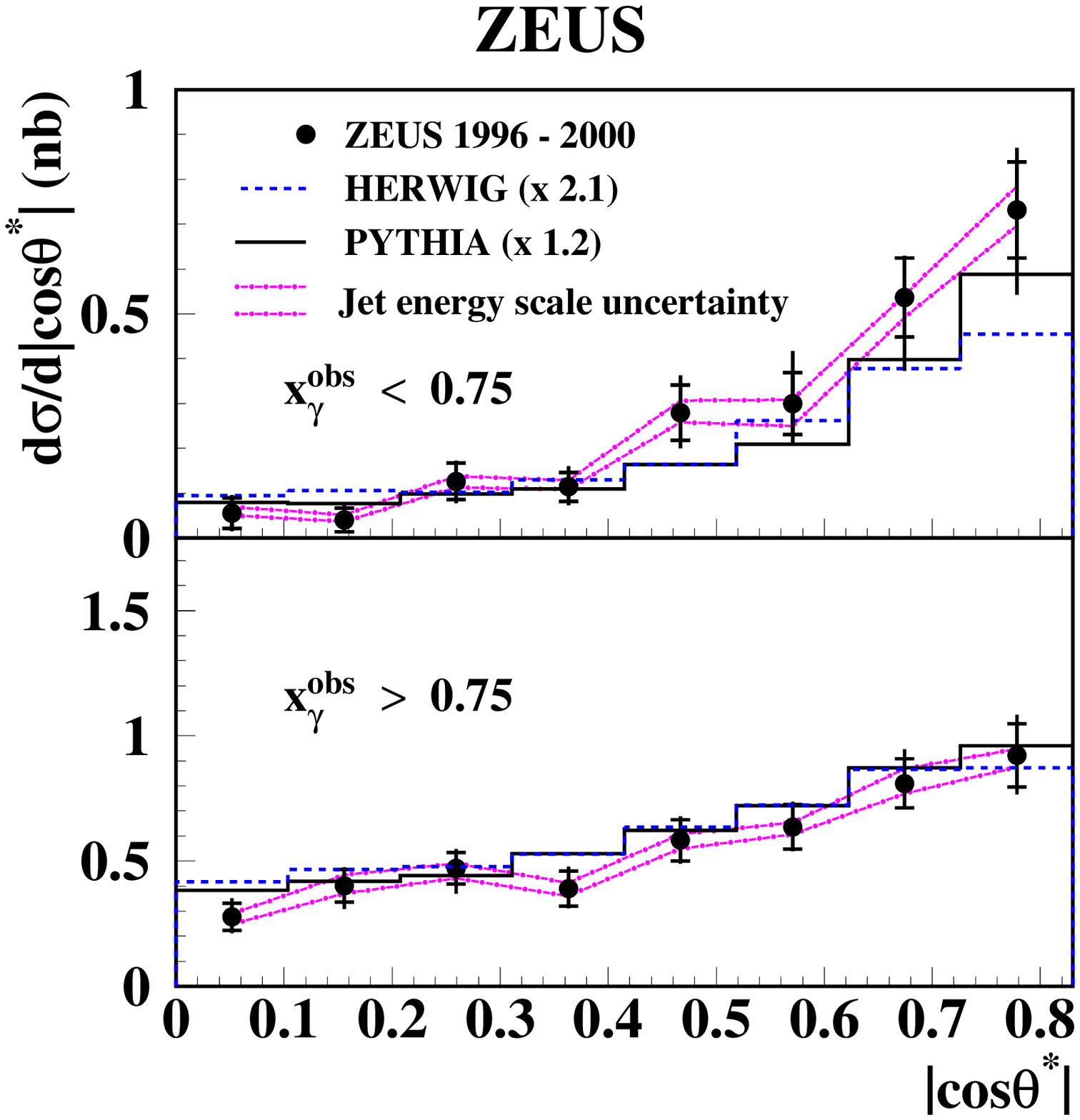,height=7.5cm}
          \end{center}          
 \vspace{-0.7cm}
\caption{Differential cross sections $d\sigma /d|\cos\theta^*|$ compared with MC simulations
for samples enriched in resolved- (upper plot) and direct- (lower plot) photon events. }
\label{fig:fig5}
       \end{figure} 
 
       \begin{figure}[h]
 \vspace{-1.3cm}
          \begin{center}
 \hspace*{-1.0cm}\psfig{figure=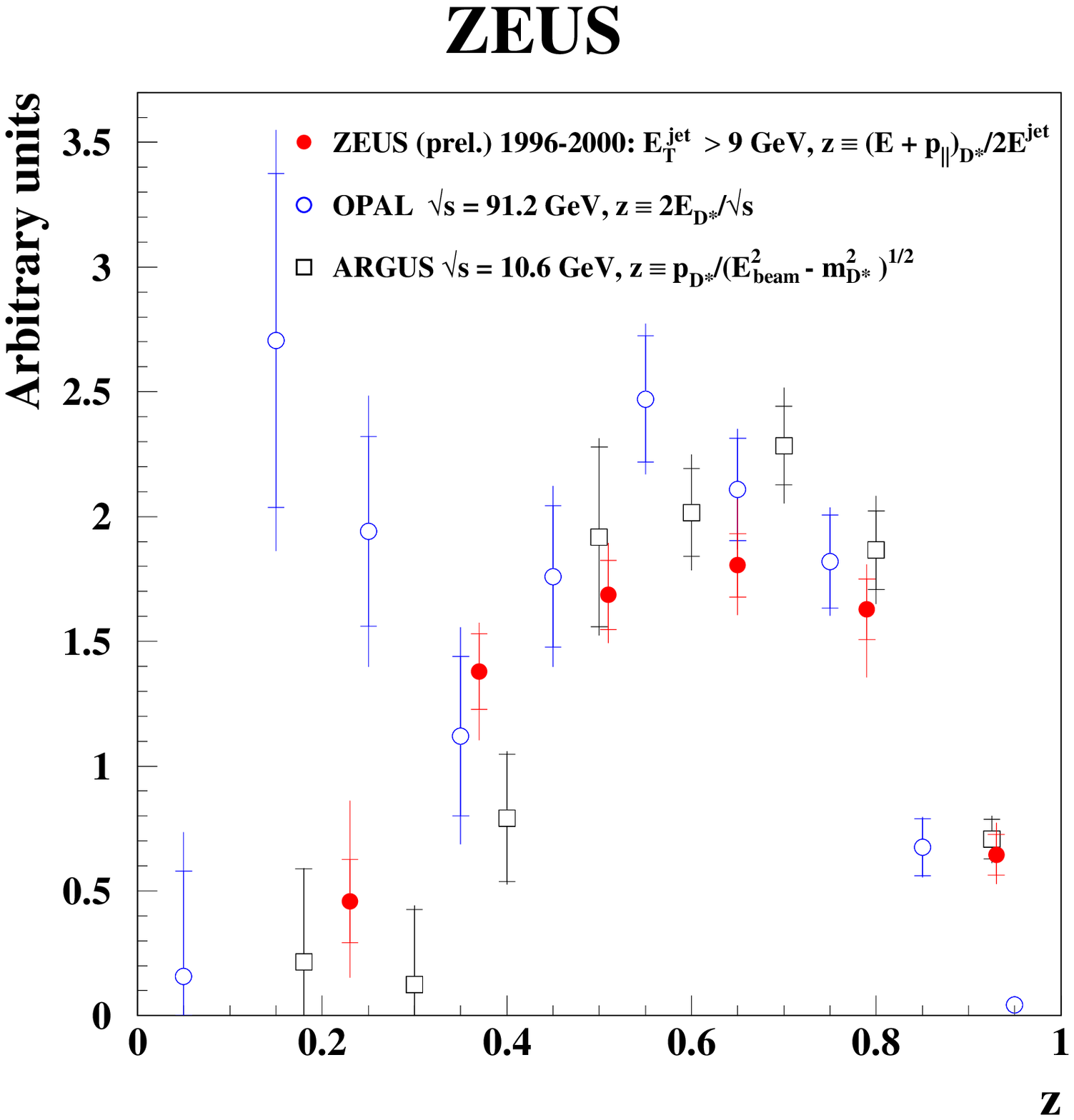,height=8.0cm}
          \end{center}          
 \vspace{-1.2cm}
\caption{Fragmentation function versus $z$ for the ZEUS data compared to $e^+ e^-$        
measurements. The data sets were normalised to 1/(bin width) for $z~>~0.3$.}
\label{fig:fig6}
       \end{figure}

The two jets can be distinguished by associating the $D^*$ meson to the closest jet in
$\eta -\phi$ space. Calling this ``$D^*$ jet" jet 1, the
 $d\sigma /d\cos\theta^*$ rise can be studied separately for the photon and proton directions.
Fig.~{7} shows the differential cross sections as a function of $cos\theta^*$ for the
resolved- and direct-enriched samples. The PYTHIA estimation of the resolved process contribution
to the direct-enriched sample (fig.{7}b) explains the asymmetric distribution in $\cos\theta^*$. 
The strong rise in 
 $d\sigma /d\cos\theta^*$ towards the photon direction for the resolved-enriched sample is    
clear evidence for charm from the photon in the LO picture.
The NLO FO predictions describe the data well for $\xgo~>~0.75$ (fig.~{7}d), but are below the data for
$\xgo~<~0.75$ (fig.~{7}c), both for the proton and photon directions. The data shapes are
reproduced. The CASCADE predictions exceed the data by $\approx 30\%$, mostly for $\xgo~>~0.75$.
Again the shapes are reasonably well reproduced.
 
\section{Charm fragmentation}
 
Charm quark hadronisation into charm mesons is parametrised by fragmentation functions,
which exist in many forms with tunable parameters fixed from fits to $e^+ e^-$ data.
A direct measurement of the fragmentation function at HERA can reduce theoretical uncertainties
and test the universality of charm fragmentation.
 
Charm-jet PHP events have been measured~\cite{ichep778} with the ZEUS detector,                       
using an integrated luminosity of $120~pb^{-1}$,
in the kinematic region $Q^2~<~1~GeV^2$,
 $130~<~W~<~280~GeV$.       At least one jet had to satisfy                                    
   $E_T^{jet}~>~9~GeV$, $|\eta^{jet}|~<~2.4$ and a $D^*$ meson with
$p_t(D^*)~>~2~GeV$, $|\eta(D^*)|~<~1.5$ had to be associated with a jet.
The fraction of jet energy carrried by the $D^*$ was defined as
            $z~=~\frac{(E+p_{\parallel})_{D^*}}{2E_{jet}}$, where                             
 $p_{\parallel}$ is the $D^*$ longitudinal momentum relative to the jet axis.
The normalised differential cross
section in $z$ was compared to the PYTHIA LO MC with the Peterson fragmentation         
function~\cite{peterson}
       $f(z)\propto~[z(1-~1/z~-~\epsilon/(1-~z))^2]^{-1}$, for various values of the
free parameter $\epsilon$. Strong sensitivity to the $\epsilon$ value was found. A fit to the
best value yielded 
$\epsilon = 0.064\pm 0.006^{+0.011}_{-0.008}$, compared to 0.053 from LO fits to the LEP data.
 
The $z$ distribution measured with the ZEUS data was compared to that from $e^+ e^-$ colliders,
where $z$ was defined as 
$z=E_{D^*}/E_{beam}$. The comparison with OPAL and ARGUS results is shown in fig.~{6}.
Similar shapes are obtained for $z~>~0.3$ with precision of the HERA data competitive with 
the LEP data. The low-$z$ peak in the OPAL data is due to gluon splitting to $c\bar c$.
The results support the universality of the charm fragmentation function.

       \begin{figure}[h]
 \vspace{-0.3cm}
          \begin{center}
          \psfig{figure=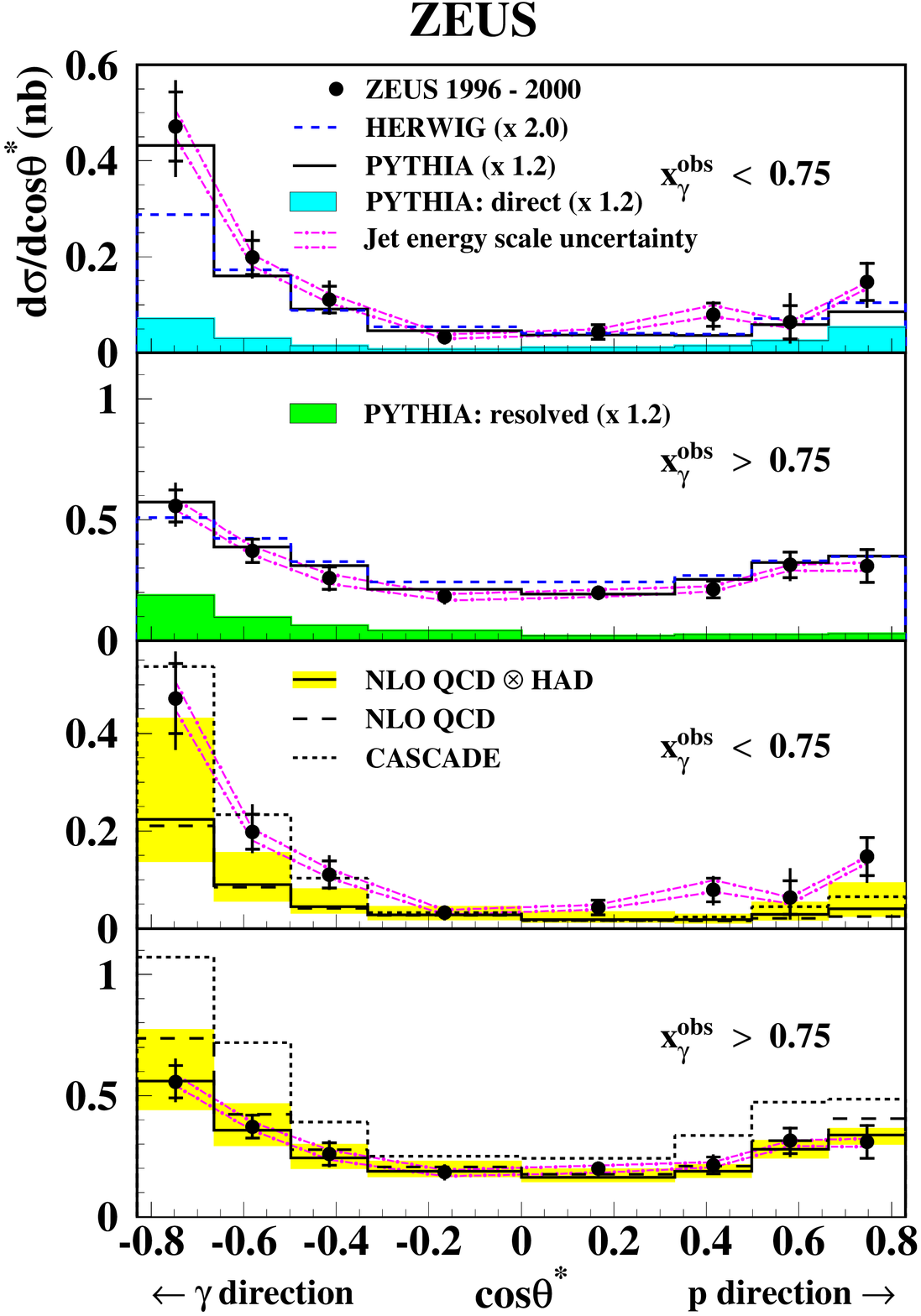,height=10.5cm}
 \put(-50,272){a)}
 \put(-50,208){b)}
 \put(-50,144){c)}
 \put(-50, 80){d)}
          \end{center}          
 \vspace{-1.2cm}
\caption{Differential cross sections $d\sigma /d\cos\theta^*$ compared with MC simulations
and NLO FO predictions
for samples enriched in resolved- (a,c) and direct- (b,d) photon events.  
The shaded areas in (a,b) are the contribution of the PYTHIA direct photon process to the 
resolved-enriched sample and of the resolved photon process to the direct-enriched sample,
respectively. The NLO uncertainty after
hadronisation is given by the shaded band. } 
\label{fig:fig7}
       \end{figure} 
 
The H1 measurements~\cite{H1DIS} of inclusive charm mesons in the DIS regime (section 2) were used
to deduce preliminary fragmentation fractions of the charm quark to the various mesons.
The results, compared with the $e^+ e^-$ world average values (in brackets), are: \\
$  f(c\to D^+)=        0.20  \pm 0.02^{+0.04+0.03 }_{-0.03-0.02 }   
 ~~~~    ( 0.23  \pm 0.02 ) $  \\
$  f(c\to D^0)=        0.66  \pm 0.05^{+0.12+0.09}_{-0.14-0.05}   
~~~~~    ( 0.55  \pm 0.03 ) $   \\
$  f(c\to D_s^+)=      0.16  \pm 0.04^{+0.04+0.05}_{-0.04-0.05}   
~~~~    ( 0.10  \pm 0.03 ) $     \\
$  f(c\to D^{*+})=        0.26  \pm 0.02^{+0.06+0.03}_{-0.04-0.02}   
~~~~    ( 0.24  \pm 0.01 ) $     \\
where the first uncertainty is statistical, the second is systematic and the third is from
theory.                  
 
In addition, the H1 measured            cross sections were used to test the isospin invariance
of the fragmentation process by calculating the ratio of neutral ($c\bar u$) to charged ($c\bar d$)
$D$-meson production \\
$ R_{u/d} =                                                                                  
 \frac{\sigma^{dir}(D^0)+\sigma(D^{*0})}{\sigma^{dir}(D^{\pm})+\sigma(D^{*\pm})}   =  
 \frac{\sigma(D^0)-\sigma(D^{*+})\times BR}{\sigma^(D^{\pm})+\sigma(D^{*\pm})\times BR}$  \\
  and to extract the strangeness suppression factor \\             
$ \gamma_s =    \frac{2 f(c\to D^+_s)}{f(c\to D^+)+f(c\to D^0)}  $                    
and the fraction  of $D$ mesons produced in a vector state                               
$P_V~=~\frac{V}{V+PS}=\frac{D^*}{D^* +D}$. Here $BR=0.677\pm 0.005$ is the
$D^{*+}\to D^0\pi^+$ branching ratio.                       The results,     compared
with previous values, are: \\
$  R_{u/d}=         1.26 \pm 0.20\pm 0.12 ~~$        (H1 prel.) \\
$~~~~~~~~~~~      1.00 \pm 0.09 ~~~~~~~~$   ($e^+ e^-$ world average) \\
$ \gamma_s =   0.36 \pm 0.10\pm 0.08~~~~~$         (H1 prel.) \\
$~~~~~~~~                0.27 \pm 0.04\pm 0.03~~~~~~$         (ZEUS)       \\
$~~~~~~~~                0.26 \pm 0.03~~~~~~~~~~~$         ($e^+ e^-$ world average) \\
$  P_V=        0.69  \pm 0.04 \pm 0.01            ~~~~~$       (H1 prel.) \\
$~~~~~~~~~        0.55   \pm 0.04\pm 0.03 $           $~~~$       (ZEUS prel.) \\
$~~~~~~~~~       0.60   \pm 0.03 ~~~~~~~~~~$          ($e^+ e^-$ world average).\\
All the HERA measurements are in good agreement with $e^+ e^-$ world average results, indicating
again the universality of charm fragmentation.

\end{document}